\documentclass[12pt]{article}
\usepackage{amsmath,amssymb}
\usepackage{amsfonts,amsthm,amsmath,amssymb,graphicx}
\usepackage{color}

 \def\ep{{\epsilon}}

 \def\frac#1#2{{#1\over #2}}

 \def\s{\sqrt}

\def\be{\begin{equation}}
\def\ee{\end{equation}}
\def\ba{\begin{eqnarray}}
\def\ea{\end{eqnarray}}

 \def\de{\partial}

 \def\f {\frac}
 \def\ti{\tilde}
 \def\ap{\alpha}

 \def\ddd{\cdot\cdot\cdot}
 \def\no{\nonumber \\}

 \def\la{\langle}
 \def\lb{\rangle}
 \def\ep{\epsilon}

\begin{document}

\begin{titlepage}
\thispagestyle{empty}

\begin{flushright}
UT-Komaba 16-4
\\
YITP-16-58
\\
IPMU16-0067
\\
\end{flushright}


\begin{center}
\noindent{{\textbf{\large Causal Evolutions of Bulk Local Excitations from CFT}}}\\
\vspace{2cm}
Kanato Goto$^{a}$, Masamichi Miyaji$^{b}$ and Tadashi Takayanagi$^{b,c}$
\vspace{1cm}

{\it
$^{a}$ Institute of Physics, The University of Tokyo\\ Komaba, Meguro-ku, Tokyo 153-8902, Japan\\
$^{b}$Yukawa Institute for Theoretical Physics,\\
Kyoto University, Kyoto 606-8502, Japan\\
$^{c}$Kavli Institute for the Physics and Mathematics of the Universe,\\
University of Tokyo, Kashiwa, Chiba 277-8582, Japan\\
}

\vskip 2em
\end{center}

\begin{abstract}
  Bulk localized excited states in an AdS spacetime can be constructed from Ishibashi states with respect to the global conformal symmetry in the dual CFT. We study boundary two point functions of primary operators in the presence of bulk localized excitations in two dimensional CFTs. From two point functions in holographic CFTs, we observe causal propagations of radiations when the mass of dual bulk scalar field is close to the BF bound. This behavior for holographic CFTs is consistent with the locality and causality in classical gravity duals. We also show that this cannot be seen in free fermion CFTs. Moreover, we find that the short distance behavior of two point functions is universal and obeys the relation which generalizes the first law of entanglement entropy.
  \end{abstract}

\end{titlepage}

\newpage

\section{Introduction}

To understand the basic mechanism of AdS/CFT correspondence \cite{Ma,GKP}, one of the most important problems is the locality of a bulk spacetime \cite{HKLL,HPPS,ADH,KL,MSZ}. One direct approach to this problem is to probe a point in the bulk AdS by locally exciting the vacuum state $|0\lb_{bulk}$ around that point. A purely CFT counterpart of this was given in \cite{MNSTW} for the AdS$_3/$CFT$_2$ setup. Consider a locally excited state $\hat{\phi}_\ap(y)|0\lb_{bulk}$ in the bulk, where $\hat{\phi}_\ap$ is a scalar field in the AdS space, dual to a primary operator $O_\ap$ and $y$ is a point in AdS. Its CFT dual $|\Psi_\ap\lb$ is obtained from a boundary state or more precisely so called Ishibashi state \cite{Is} for the global conformal symmetry (see also \cite{Ve} for a slightly different proposal using the full Virasoro symmetry). This precisely reproduces the bulk two point function $\la \hat{\phi}_\ap(y_1)\hat{\phi}_\ap(y_2)\lb$ assuming that the bulk scalar field is a free massive scalar field \cite{MNSTW,NO}. Indeed in the section two of the present paper, we will prove that the state $|\Psi_\ap\lb$ in \cite{MNSTW} is precisely the same as that obtained by acting the local bulk field operator in \cite{HKLL} on the vacuum. This state corresponds to the large $N$ limit (or equally large $c$ limit) and $1/N$ corrections modify the identification of the dual state \cite{NO,NaOo}. We can also generalize this construction of locally excited states in higher dimensions \cite{NO,Wang}.

The locality in AdS/CFT requires the large $N$ limit with a large gap in the spectrum of primary fields in a CFT \cite{HPPS,Ha,FKW,HKS}, called holographic CFTs. Therefore it is intriguing to see how this condition appears when we regard $|\Psi_\ap\lb$ as a locally excited state in the AdS. However, the computation of the bulk two point function is universal in any CFTs and the result reproduces the expected gravity result in AdS even for non holographic CFTs such as free CFTs. Therefore we need to look at different quantities which depend on more details of CFTs.

Motivated by this, the main purpose of this paper is to study the CFT two point function of an primary field $O(x)$ for the exited state $|\Psi_\ap\lb$, which is written as a four point function
$\la \Psi_\ap|O(x)O(y)|\Psi_\ap\lb$ for the CFT vacuum. When the conformal dimension of $O(x)$ is large (but not too large), we can evaluate this two point function from the geodesic length in the spacetime dual to the excited state $|\Psi_\ap\lb$. Thus this quantity is also directly related to the holographic entanglement entropy \cite{RT}. Therefore this quantity is directly related to the bulk geometry and we will use this to study the time evolution of bulk geometry, which strongly depends on whether a CFT is holographic or not. If the bulk theory is local and causal, then the excitations of light fields will spread relativistically inside a light cone. Indeed we will observe such an evolution for holographic CFTs when the conformal dimension of $O_\ap$ is close to the Breitenlohner-Freedman (BF) bound \cite{BF}. On the other hand, we find that this does not happen for a free fermions CFT.

This paper is organized as follows: In section two, after we briefly review the state $|\Psi_\ap\lb$ dual to a localized excitation in the bulk AdS, we will show that this state can also be derived from the known CFT dual of bulk local field. In section three, we formulate the computation of the two point function for the excited state in terms of four point function of primary fields. In section four, we focus on a large $c$ CFT (i.e. holographic CFT) and evaluate the two point function. We will show that our result agrees with our expectation from AdS/CFT.  In section five, we study a different parameter region of conformal dimensions in a large $c$ CFT.  In the appendix A, we compute the two point function in a free fermion CFT.

\section{CFT Duals of Bulk Localized States}

\subsection{Construction}

In \cite{MNSTW}, CFT states which are dual to bulk localized states in AdS$_3/$CFT$_2$ are introduced in terms of specific descendants of the global part (i.e. $SL(2,R)^2$) of Virasoro symmetry. At $t=0$, this state is obtained by exciting only a point in the time slice of AdS$_3$ (i.e. $H_2$). If we set this point to the origin $r=0$ of global AdS$_3$
\be
ds^2=-(r^2+R^2)dt^2+\f{R^2dr^2}{r^2+R^2}+r^2d\theta^2,  \label{gads}
\ee
the dual state should be invariant under all $SL(2,R)^2$ transformations which keep the point invariant.　This leads to the following constraints
\be
(L_0-\ti{L}_0)|\Psi_\ap\lb=(L_{\pm 1}+\ti{L}_{\mp 1})|\Psi_\ap\lb=0. \label{constrfg}
\ee
The solutions to (\ref{constrfg}) can be constructed from (analogue of) Ishibashi states $|J_\ap\lb$ for the $SL(2,R)^2$ algebra for various primaries and are explicitly given as follows:
\be
|\Psi_\ap\lb=\s{\cal{N}}\sum_{k=0}^\infty (-1)^k e^{-\ep k}|k,\ap\lb, \label{ishir}
\ee
where we defined an orthonormal basis:
\be
|k,\ap\lb=\f{1}{N_k}(L_{-1}\ti{L}_{-1})^k|\ap\lb,
\ee
where $|\ap\lb$ is a primary state with the chiral and anti chiral conformal dimension $h_\ap=\bar{h}_\ap$; we also define the total conformal dimension by $\Delta_\ap=h_\ap+\bar{h}_\ap(=2h_\ap)$. $N_k$ is defined as
\be
N_k=\f{\Gamma(k+1)\Gamma(2h_\ap+k)}{\Gamma(2h_\ap)}.  \label{ishik}
\ee
We introduced a parameter $\ep$ in (\ref{ishir}) which provides a UV regularization and is taken such that
$\f{1}{c}\ll \ep\ll 1$ in order to make its holographic meaning clear, where $c$ is the central charge of the CFT. The overall normalization which guarantees $\la \Psi_\ap|\Psi_\ap\lb=1$, is given by
\be
{\cal N}=\f{1}{\sum_{k=0}^\infty e^{-2k\ep}}=1-e^{-2\ep}. \label{nork}
\ee
It is useful to evaluate the averaged energy (or conformal dimension) $\Delta_{\Psi_\ap}$ of $|\Psi_\ap\lb$ as
\be
\Delta_{\Psi_\ap}=\la \Psi_\ap|L_0+\ti{L}_0|\Psi_\ap\lb=(1-e^{-2\ep})\sum_{k=0}^\infty
(2k+\Delta)e^{-2\ep k}=\Delta_\ap+\f{2}{e^{2\ep}-1}\simeq \Delta_\ap+\f{1}{\ep}.  \label{amd}
\ee

We will also consider the time evolution\footnote{Here we omit the $t$ dependent phase factor
$e^{ict/12}$ as it does not physically affect our arguments}
 of this state (\ref{ishir}):
\be
|\Psi_\ap(t)\lb=e^{-iHt}|\Psi_\ap\lb,
\ee
where $H=L_0+\bar{L}_0$. It is obvious that this state follows the periodicity
\be
|\Psi_\ap(t+\pi)\lb=|\Psi_\ap(t)\lb.
\ee
Also at $t=\pi/2$, the state coincides with the Ishibashi state
\be
|\Psi_\ap(\pi/2)\lb=|J_\ap\lb=\s{\cal{N}}\sum_{k=0}^\infty e^{-\ep k}|k,\ap\lb.
\label{Ishi}
\ee

\subsection{Relation to HKLL Prescription}
Next we will show that this state $|\Psi_\ap\lb$ can also be obtained from
the CFT dual of bulk local field constructed in \cite{HKLL} (so called HKLL construction).
We write the primary field corresponding to the primary state $|\ap\lb$ as $O_\ap$.

We introduce a coordinate of global AdS$_3$ as follows
\be
ds^2=\f{R^2}{\cos^2\eta}(-d\tau^2+d\eta^2+\sin^2\eta d\theta^2).
\ee
We can also use the coordinate $\rho$ defined by
\be
\sinh\rho=\f{\sin\eta}{\cos\eta},\ \ \ \cosh\rho=\f{1}{\cos\eta},
\ee
so that the metric looks like
\be
ds^2=R^2(-\cosh^2\rho d\tau^2+d\rho^2+\sinh^2\rho d\theta^2).
\ee

The bulk local field operator $\hat{\phi}_\ap(\tau,\eta,\theta)$ is expressed
in terms of the dual local operator $O_\ap(\tau,\theta)$ in the CFT as
\be
\hat{\phi}_\ap(\tau,\eta,\theta)=\int^{\pi/2}_{-\pi/2}d\tau'
\int^{2\pi}_0 d\theta' K(\tau',\theta'| \tau,\eta,\theta) O_\ap(\tau',\theta').
\ee

When $\eta=0$, $\tau=0$ (i.e. the origin of global AdS$_3$), we find the expression \cite{HKLL}:
\be
K(\tau',\theta'| 0,0,\theta)
=\f{1}{2\pi^2}e^{i\Delta_\ap \tau'} {{}_2F_{1}}(1,1,\Delta_\ap,-e^{2i\tau'}),
\ee
where $\Delta_\ap(=2h_\ap)$ is the total conformal dimension of the primary operator $O$.
Note that the above hypergeometric function is expressed as
\be
{{}_2F_{1}}(1,1,\Delta_\ap,-e^{2i\tau})=\Gamma(\Delta_\ap)\sum_{n=0}^\infty \f{\Gamma(n+1)}{\Gamma(n+\Delta_\ap)}(-1)^n e^{2in\tau}.
\ee

In general a primary field $O$ is represented as the following mode expansion (see e.g. \cite{cftbook}):
\be
O(\tau,\theta)=\sum_{n,m\in Z}O_{n-h_\ap,m-h_{\ap}}e^{i(n-h_\ap)(\tau+\theta)}e^{i(m-h_\ap)(\tau-\theta)}.
\ee

Therefore, by acting on the vacuum, the bulk state $\hat{\phi}_\ap(\tau,0,\theta)|0\lb$  is dual to the CFT state given by (after the integration of $\tau'$ and $\theta'$)
\be
\hat{\phi}_\ap(0,0,\theta)|0\lb=\sum^\infty_{n=0}\f{\Gamma(n+1)\Gamma(2h_\ap)}{\Gamma(n+2h_\ap)}(-1)^n
\cdot O_{-n-h_\ap,-n-h_\ap}|0\lb.
\ee

On the other hand, using the commutation relation
\ba
&&[L_m,O_{n-h_\ap,k-h_\ap}]=\left(m(h_{\ap}-1)-n+h_{\ap}\right)O_{n+m-h_{\ap},k-h_\ap},\no
&&[\ti{L}_m,O_{n-h_\ap,k-h_\ap}]=\left(m(h_{\ap}-1)-k+h_{\ap}\right)O_{n-h_{\ap},k+m-h_\ap},
\ea
we obtain the following identifications:
\be
(L_{-1})^k(L_{-1})^l|\ap\lb=\Gamma(k+1)\Gamma(l+1)\cdot O_{-k-h_{\ap},-l-h_{\ap}}|0\lb, \ \  \ \ \ (k,l=0,1,2,\ddd).
\ee
By using these relations we can express the state $|\Psi_\ap\lb$ as
\be
|\Psi_\ap\lb=\sum^\infty_{n=0}\f{\Gamma(n+1)\Gamma(2h_{\ap})}{\Gamma(n+2h_{\ap})}(-1)^n
\cdot O_{-n-h_{\ap},-n-h_{\ap}}|0\lb.
\ee

Thus $|\Psi_\ap\lb$ and $\hat{\phi}(0,0,\theta)|0\lb$ are equivalent.

\section{Two Point Functions for $|\Psi_\ap\lb$: Formulation}

The main quantities we study in this paper are two point functions of primary operators $O$
in the presence of the excitation given by the state $|\Psi_\ap\lb$:
\be
\la OO\lb_{\Psi_\ap}=\la \Psi_\ap|OO|\Psi_\ap\lb.  \label{twopap}
\ee

\subsection{Conventions of Four-point Functions}

To evaluate (\ref{twopap}), we need to know the four-point functions of the form $\la O_\ap O_\ap O O \lb$. Therefore we would like to start with a brief summary of properties of four-point functions and our conventions.

We focus on the four point function of the form
\be
\la O_A(x)O_A(y)O_B(z)O_B(w)\lb,  \label{fpt}
\ee
where the (chiral) conformal dimensions of $O_A$ and $O_B$ are denoted by $h_A$ and $h_B$.

By introducing the cross ratio
\be
\eta=\f{(x-y)(z-w)}{(x-z)(y-w)},
\ee
we can express the four point function (\ref{fpt}) as follows
\ba
\la O_A(x)O_A(y)O_B(z)O_B(w)\lb &=&
|x-z|^{-4h_{A}}|y-w|^{-4h_{B}}|z-w|^{4h_{A}-4h_{B}}\cdot |F(\eta)|^2 \no
&=& |x-y|^{-4h_{A}}|z-w|^{-4h_{B}}\cdot |\eta|^{4h_{A}}\cdot |F(\eta)|^2.
\ea
Note that actually we need to take the summation $\Sigma_a |F_a(\eta)|^2$ over conformal
blocks. However in the examples we discuss below only essentially involves only one conformal block.

This function $F(\eta)$ is equal to the following limit of the four point function:
\be
\lim_{w\to\infty}|w|^{4h_{B}}\cdot \la O_A(0)O_A(\eta)O_B(1)O_B(w)\lb=F(\eta).
\ee

\subsection{Example 1: 2D Free Dirac Fermion}

Here we summarize the result of four point function in a free Dirac fermion, which will be employed later. It is useful to note that via the bosonization, a massless free Dirac fermion is equivalent to a massless free scalar.
We choose
\be
O_A=e^{i\phi}=\psi,\ \ \ O_B=e^{i\beta\phi},
\ee
and consider $\la O_A(x)\bar{O}_A(y)O_B(z)\bar{O}_B(w)\lb$, which can be treated as in the previous way. Assume the OPE $\phi(z)\phi(w)\sim -\log (z-w)$. Note that $h_{A}=1/2$ and
$h_{B}=\beta^2/2$.

In this example, we can explicitly compute the four point function and find
\ba
\la O_A(x)\bar{O}_A(y)O_B(z)\bar{O}_B(w)\lb
=\left|\f{1}{(x-y)(z-w)^{\beta^2}(1-\eta)^\beta}\right|^2.  \label{fermr}
\ea
In other words, we have
\be
\eta^{2h_{A}}F(\eta)=\f{1}{(1-\eta)^\beta}.  \label{dedf}
\ee

Even though we do not expect any classical gravity dual for a free fermion CFT, this provides us a toy example which allows full analytical calculations. We will summarize the result of two point functions for our excited state $|\Psi_\ap\lb$ in the appendix $A$.

\subsection{Example 2: Holographic CFT$_2$ with $h_A=O(1)$ and $h_B=O(c)$}

In AdS$_3/$CFT$_2$, a CFT with its classical gravity dual is characterized by the following properties \cite{HPPS,Ha,FKW}:
(1) it has a large central charge $c$ and (2) the low energy spectrum is sparse.
For such CFTs (called holographic CFTs), when we assume the conformal dimensions $h_A=O(1)$ and $h_B=O(c)$,
the function $F(\eta)$ is found as follows by using the result in \cite{FKW,FKWt}:
\be
F(\eta)=\left(\f{\ap_B}{1-(1-\eta)^{\ap_B}}\right)^{2h_{A}}\cdot (1-\eta)^{-h_{A}(1-\ap_B)},
\label{lcfg}
\ee
where $\ap_B=\s{1-24h_{B}/c}$.

If we expand in terms of small $\eta$, we have
\be
F(\eta)=\eta^{-2h_{A}}\left(1+\f{2h_{A}h_{B}}{c}\eta^2+\ddd\right)  \label{lcf}
\ee

\subsection{Example 3: Holographic CFT$_2$ with $h_A/c\ll 1$ and $h_B/c\ll 1$}

Consider a holographic CFT in two dimension when the conformal dimensions take their values in the range:
\be
\f{h_A}{c}\ll 1, \ \ \ \f{h_B}{c}\ll 1, \ \ \ \f{h_A h_B}{c}=\mbox{fixed}.
\ee

In this case by using eq.(B.34) in \cite{FKW}, we find
\be
F(\eta)=\eta^{-2h_A}\cdot e^{\f{2h_A h_B}{c}{}_2F_1(2,2,4,\eta)}, \label{hh}
\ee
where note that
\be
{}_2F_1(2,2,4,\eta)=\f{6}{\eta}\left(\eta\log(1-\eta)-2\log(1-\eta)-2\eta\right).
\ee
We can confirm that the leading linear term when we Taylor expand the exponential in (\ref{hh}) agrees with
the leading linear term in the expansion w.r.t $h_B/c$ of (\ref{lcfg}).

\subsection{Computations of Two Point Functions for $|\Psi_\ap\lb$}

We now set the two primary operators as $O_A=O_{\ap}$ and $O_B=O$ in (\ref{fpt}). Thus we have $h_{A}=h_\ap$ and $h_{B}=h_O$. Using this, we would like to compute the two point function $\la \Psi_\ap|O(1)O(e^{-i\sigma})|\Psi_\ap\lb$ on a cylinder for the excited state $|\Psi_\ap\lb$. The coordinate $\sigma$ corresponds to the circular direction with $2\pi$ periodicity at fixed time $t=0$.
The primary operator $O(x)$ is expected to probe the physical properties of the excited state $|\Psi_\ap\lb$.

We can identify $L_{-1}$ and $\ti{L}_{-1}$ as the derivative $\de_x$ and $\ti{\de}_x$ which act on
the four point function (\ref{fpt}) in the limit $x\to 0$ and $u=1/y\to 0$. Thus we can express this two point function as the derivatives of four point function. More explicitly, taking the time evolution into account, we have
\ba
&& \la \Psi_\ap(t)|O(1)O(e^{-i\sigma})|\Psi_\ap(t)\lb \no
&& ={\cal N}\cdot |1-e^{-i\sigma}|^{-4h_O}\cdot\lim_{x\to 0, u\to 0} {\cal D}_x{\cal D}_u \left[|1-xu|^{-4h_\ap}|\eta|^{4h_\ap}|F(\eta)|^2\right],\ \ \
\label{jwje}
\ea
where ${\cal N}$ is the normalization factor (\ref{nork}). The operations ${\cal D}_x$ and ${\cal D}_u$ are defined by
\ba
&&{\cal D}_x=\sum_{p=0}^\infty \f{ e^{-ip(2t+\pi)}\cdot e^{-p\ep}}{N_p}(\de_x)^p(\bar{\de}_x)^p, \label{ddxu} \\
&&{\cal D}_u=\sum_{q=0}^\infty \f{ e^{iq(2t+\pi)}\cdot e^{-q\ep}}{N_q}(\de_u)^q(\bar{\de}_u)^q,\label{ddxuu}
\ea
where the coefficient $N_p$ is defined in (\ref{ishik}). Also the cross ratio $\eta$ is given by
\ba
&& \eta=\f{(e^{i\sigma}-1)(xu-1)}{(e^{i\sigma}-u)(x-1)}
=e^{i\sigma}\f{(1-e^{-i\sigma})(xu-1)}{(e^{i\sigma}-u)(x-1)}, \no
&& 1-\eta=\f{(1-u)(1-e^{i\sigma}x)}{(e^{i\sigma}-u)(1-x)}=e^{-i\sigma}
\cdot\f{(1-u)(1-e^{i\sigma}x)}{(1-e^{-i\sigma}u)(1-x)} . \label{kkh}
\ea

It is a simple exercise to confirm the correct normalization when $O$ is the identity operator
by setting $|\eta|^{4\Delta_\ap}|F(\eta)|^2=1$:
\be
\la \Psi_\ap|\Psi_\ap\lb={\cal N}\cdot \lim_{x\to 0, u\to 0} {\cal D}_x{\cal D}_u \left[|1-xu|^{-4\Delta_\ap}\right]
={\cal N}\cdot\sum_{k=0}^\infty e^{-2\ep k}=1.
\ee

In general CFTs, it is not easy to take the infinite summation (\ref{ddxu}) and (\ref{ddxuu}).
Indeed, in examples we will discuss later, we often approximate them by replacing the infinite summation by a finite sum $\sum^{n}_{p=0}$ and $\sum^{n}_{q=0}$. We call this an order $n$ approximation. It is obvious that if we want reliable results for smaller $\epsilon$, we need to increase $n$ accordingly such that $n\sim 1/\ep$.

\section{Two Point Functions for $|\Psi_\ap\lb$: Holographic CFTs}

Here we study two point functions of primary operator $O(x)$ for the excited state $|\Psi_\ap\lb$ in holographic CFTs to the leading order of $1/c$ expansion.  In this section we assume that the conformal dimension of $O(x)$ is large enough to be order $c$ and this amplifies the time dependence of the two point functions. Here we should note that  $|\Psi_\ap\lb$ was originally constructed for the pure AdS$_3$. To be exact, in the presence of heavy operator $O(x)$, some corrections are needed for the precise form of $|\Psi_\ap\lb$. However, as long as the bulk excitations created by $|\Psi_\ap\lb$ do not significantly overlap with those created by $O(x)$, we expect that we can neglect such corrections. Indeed our final results qualitatively confirm this expectation.

\subsection{Behavior near $\sigma=0$ (small $\eta$ expansion)}

If we assume $\eta$ is very small or equally $\sigma$ is small in (\ref{kkh}) , the two point function is expanded as follows
\ba
&& \la \Psi_\ap|O(1)O(e^{-i\sigma})|\Psi_\ap\lb \no
&& ={\cal N}\cdot |1-e^{-i\sigma}|^{-4h_O}\cdot
\lim_{x, u\to 0} {\cal D}_x{\cal D}_u\left[|1-xu|^{-4h_\ap}\left(1+\f{2h_\ap h_O}{c}\eta^2
+\f{2h_\ap h_O}{c}\bar{\eta}^2\right)\right]. \label{etaex}
\ea
It is already clear from the above expression that only diagonal summations i.e. $(\de_x\de_u\bar{\de}_x\bar{\de}_u)^k$ do contribute and therefore the phase factors $e^{i\pi (q-p)}$ in ${\cal D}_x$ and ${\cal D}_u$ do not contribute. This means that the above terms are invariant under the time evolution.

\subsubsection{$h_\ap=1$ Case}

Let us focus on the special case $h_\ap=1$ for simplicity and evaluate the two point function up to $O(\sigma^2)$ in the $\sigma\to 0$ limit. In this case we obtain
\ba
&& \la \Psi_\ap|O(1)O(e^{-i\sigma})|\Psi_\ap\lb\cdot |1-e^{-i\sigma}|^{4h_O} \no
&& =1+{\cal N}\cdot\f{2h_O}{c}\cdot\lim_{x, u\to 0} {\cal D}_x{\cal D}_u\left[\f{(e^{i\sigma}-1)^2}{(1-\bar{x}\bar{u})^{2}(e^{i\sigma}-u)^{2}(1-x)^{2}}+(h.c.)\right]\no
&& =1+{\cal N}\cdot\f{2h_O}{c}\cdot\sum_{k=0}^\infty (k+1)(1-e^{-i\sigma})^2e^{-ik\sigma-2\ep k}+(h.c.)\no
&&=1+\f{2h_O}{c}(1-e^{-2\ep})\left(\f{(1-e^{-i\sigma})^2}{(1-e^{-i\sigma-2\ep})^2}
+\f{(1-e^{i\sigma})^2}{(1-e^{i\sigma-2\ep})^2}\right).
\label{firstl}
\ea

We should note that this result is reliable only when the second
term (perturbative correction) is very small. When $\sigma\ll 1$ , we find
\be
\la \Psi_\ap|O(1)O(e^{-i\sigma})|\Psi_\ap\lb\cdot |1-e^{-i\sigma}|^{4h_O}\simeq 1-\f{4h_O\cdot\sigma^2}{c(1-e^{-2\ep})}. \label{firstjl}
\ee
Using the averaged energy (\ref{amd}), we can rewrite this as follows
\be
\la \Psi_\ap|O(1)O(e^{-i\sigma})|\Psi_\ap\lb\cdot |1-e^{-i\sigma}|^{4h_O}\simeq 1-\f{\Delta_{\Psi_\ap}\Delta_O\cdot\sigma^2}{c}, \label{firstjg}
\ee
where we used $2h_\ap=\Delta_\ap=2$.

This short distance behavior (\ref{firstjl}) is analogous to that of entanglement entropy, called the first law like relation \cite{BNTU} (see also \cite{ABS}). We take the operator $O$ to be a twist operator for the replica method and its conformal dimension is given by $h_O=\f{c}{24}(n-1/n)$. In this case, by taking a derivative of $n$ setting $n=1$ we obtain the increased amount of entanglement from (\ref{firstjg})
as follows:
\be
\Delta S_A\simeq \f{\Delta_{\Psi_\ap}\sigma^2}{6}.
\ee
This agrees with the prediction from the first law like relation in \cite{BNTU}.
In section (\ref{sbtz}) we will confirm this behavior (\ref{firstjg}) from holographic calculations of a BTZ black hole.

\subsubsection{General $h_\ap$}

For more general $h_{\ap}$ we find the following expression when $\sigma\ll 1$
\ba
&& \la \Psi_\ap|O(1)O(e^{-i\sigma})|\Psi_\ap\lb\cdot |1-e^{-i\sigma}|^{4h_O} \no
&& =1-{\cal N}\cdot \f{4h_{\ap}h_O}{c}\cdot A\cdot \sigma^2,
\ea
where $A$ is computed as
\ba
A=\sum_{k=0}^\infty \sum_{p=0}^k \left[\f{k!\cdot (k-p+1)^2\cdot (p+2h_{\ap}-3)(p+2h_{\ap}-4)\ddd
(2h_{\ap}-2)}{p!\cdot (k+2h_{\ap}-1)(k+2h_{\ap}-2)\ddd
(2h_{\ap})} e^{-2\ep k}\right].
\ea
Indeed we can confirm
\be
A=\f{1}{1-e^{-2\ep}}\left[1+\f{1}{h_\ap(e^{2\ep}-1)}\right],  \label{ght}
\ee
which leads to the formula (\ref{firstjg}).

\subsubsection{Time-dependence near $\sigma=0$ in a one more higher order}

In the perturbation around $\eta=0$, the term which has non-trivial time dependence appears from the order $O(\eta^2\bar{\eta}^2)$. More explicitly, we can replace the parenthesis in (\ref{etaex}) with
\be
\left(1+\f{2h_\ap h_O}{c}\eta^2
+\f{2h_\ap h_O}{c}\bar{\eta}^2+4\left(\f{h_\ap h_O}{c}\right)^2\eta^2\bar{\eta}^2\right).
\ee
Assuming $h_{\ap}=1$ we get
\ba
&& \la \Psi_\ap|O(1)O(e^{-i\sigma})|\Psi_\ap\lb\cdot |1-e^{-i\sigma}|^{4h_O} \no
&& = (\mbox{eq.\ref{firstl}})+4{\cal N}\left(\f{h_\ap h_O}{c}\right)^2|1-e^{-i\sigma}|^4\cdot I\no
&&\simeq (\mbox{eq.\ref{firstl}})+4{\cal N}\left(\f{h_O}{c}\right)^2\sigma^4\cdot G,
\ea
where $G$ at time $t$ is computed as
\ba
&& G\equiv
\lim_{x, u\to 0} {\cal D}_x{\cal D}_u\left[\f{1}{|1-e^{-i\sigma}u|^4|1-x|^4}\right] \no
&& =\f{1}{(1+e^{it-\ep})^2(1+e^{-it-\ep})^2}.
\ea
Thus we can conclude that under the time evolution, the two point function
does not change at the order $O(\sigma^2)$, but increases at the order $O(\sigma^4)$.

\subsection{Two Point Functions from BTZ} \label{sbtz}

Before we present our main results, it is useful to study a holographic calculation of two point functions for an excited state dual to a BTZ black hole \cite{BTZ}
\be
ds^2=(r^2+R^2-M)d\tau^2+\f{R^2 dr^2}{R^2+r^2-M}+r^2d\theta^2,
\ee
where the mass parameter $M$ is proportional to the mass $m$ of the object which constitutes the black hole:
\be
m=\f{M}{8G_NR^2}.
\ee
The mass $m$ is related to the conformal factor via the familiar relation \cite{GKP}
\be
\Delta=-1+\s{m^2R^2+1}.
\ee

The energy $\Delta_{\Psi_\ap}$ of the state $|\Psi_\ap\lb$ is given by (\ref{amd}) and we can identify \be
m_{\Psi_\ap}R\simeq \Delta_{\Psi_\ap}.
\ee

The two point function in AdS space is generally approximated as
\be
\la O(x)O(y)\lb\sim e^{-m_O L},  \label{georg}
\ee
where $m_O$ is the mass of the field dual to operator $O$ given by $m_OR\simeq \Delta_O$
and $L$ is the geodesic distance of the two points in AdS.

The geodesic distance $L$, which connects two points at the AdS boundary: $(r_\infty,\theta_\infty)$ and $(r_\infty,-\theta_\infty)$ at $\tau=0$ (note $r_{\infty}\to \infty$), is given by
\ba
&& L=2R\log\f{r_\infty}{\s{R^2-M+r_*^2}} ,\no
&& \cos\left(\f{2\s{R^2-M}}{R}\theta_\infty\right)=\f{r_*^2-R^2+M}{r_*^2+R^2-M}. \label{btzgo}
\ea

Finally we get
\be
L=2R\cdot \log\f{\s{2}r_\infty\s{1-\cos\left(2\theta_\infty\s{1-12mR/c}\right)}}{R\s{1-12mR/c}}. \label{btzgh}
\ee

Now we set $\sigma=2\theta_\infty$ and then we can estimate assuming $\sigma\ll 1$ as follows:
\be
L-L_0\simeq \f{mR^2}{c}\sigma^2.
\ee
This leads to
\be
\f{\la\Psi_\ap|O(1)O(e^{-i\sigma})|\Psi_\ap\lb}{\la0|O(1)O(e^{-i\sigma})|0\lb}\simeq 1-\f{(m_OR)(mR)}{c}\sigma^2+\ddd=1-\f{\Delta_{\Psi_\ap}\Delta_O}{c}\sigma^2+\ddd.
\ee
This behavior agrees with (\ref{firstjg}).

However, we should note that we cannot always approximate our excited state $|\Psi_\ap(t)\lb$ by a BTZ solution, though they have the same energy. This is simply because the BTZ solution is static, while our excited state is time-dependent (except the near AdS boundary region). Nevertheless, if we take the large $\ep$ limit, then  $|\Psi_\ap(t)\lb$ is reduced to the primary state $|\ap\lb$ which is static and is known to be described by a BTZ black hole, using the result in \cite{FKW}. When $\epsilon$ is not large, we expect a non-trivial time evolution with a periodicity $\pi$, whose study is the main issue of the rest of this paper.

\subsection{Full Numerical Calculation}

To understand the behavior away from $\sigma=0$ we need a numerical calculation.
For this we approximate the infinite summation in (\ref{ddxu}) and (\ref{ddxuu}) by a finite sum
such that $0\leq p,q\leq n$ (i.e. order $n$ approximation). We can always confirm the validity of this approximation by comparing the small $\sigma$ behavior with the universal exact result (\ref{firstjg}).

Let us choose $h_{\ap}=1/2$ for example\footnote{Actually $h_{\ap}=1/2$ has a special meaning in the dual gravity side. It corresponds to the mass $m^2=-1/R^2$ which saturates the Breitenlohner-Freedman (BF) bound \cite{BF}. In other words, it corresponds to the lightest scalar field in AdS.} and take $h_{O}$ to be $O(c)$ with $\ap_O=\s{1-\f{24h_{O}}{c}}$ so that we can employ
the formula (\ref{lcfg}). The full expression is given by the following formula

\ba
&& \f{\la \Psi_\ap|O(1)O(e^{-i\sigma})|\Psi_\ap\lb}{\la 0|O(1)O(e^{-i\sigma})|0\lb} \no
&& =\la \Psi_\ap|O(1)O(e^{-i\sigma})|\Psi_\ap\lb\cdot |1-e^{-i\sigma}|^{4h_{O}} \no
&& ={\cal N}\cdot |1-e^{-i\sigma}|^2\cdot
\lim_{x, u\to 0} {\cal D}_x{\cal D}_u K, \label{tworat}
\ea
where $K$ is given by
\be
K=\f{\ap^2\left[|1-u|\cdot|1-e^{-i\sigma}u|\cdot|1-x|\cdot|1-e^{i\sigma}x|\right]^{\ap-1}}
{\left|(1-e^{-i\sigma}u)^\ap (1-x)^\ap -e^{-i\ap\sigma}(1-u)^\ap(1-e^{i\sigma}x)^\ap\right|^2}.
\ee

In particular, if we take the limit $\ap_O=0$ or equally $h_{O}=\f{c}{24}$, the expression is simplified as follows
\ba
K=|(1-u)(1-x)(1-e^{-i\sigma}u)(1-e^{i\sigma}x)|^{-1}\cdot
\left|-i\sigma+\log \f{(1-u)(1-e^{i\sigma}x)}{(1-e^{-i\sigma}u)(1-x))}\right|^{-2}.\no
\ea

\begin{figure}[ttt]
   \begin{center}
     \includegraphics[height=4cm]{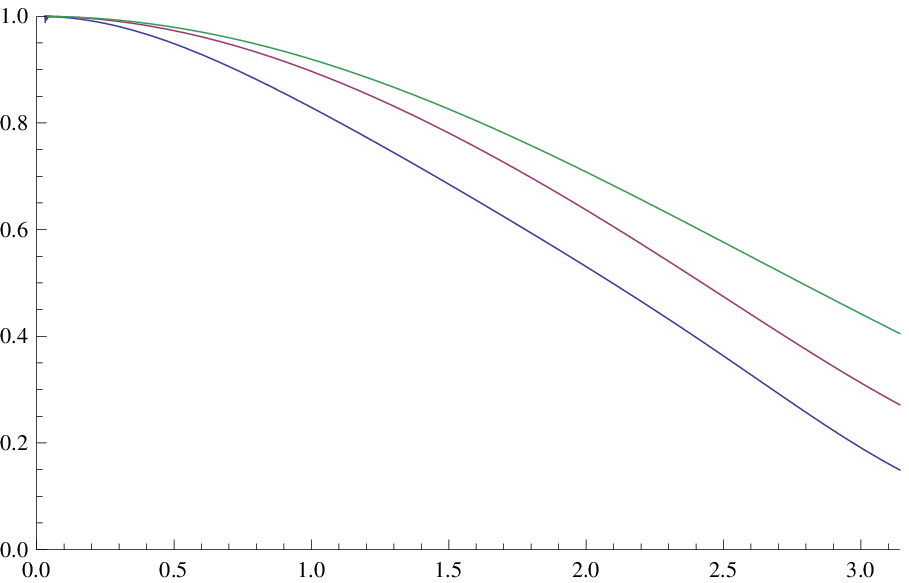}
     \hspace{1cm}
      \includegraphics[height=4cm]{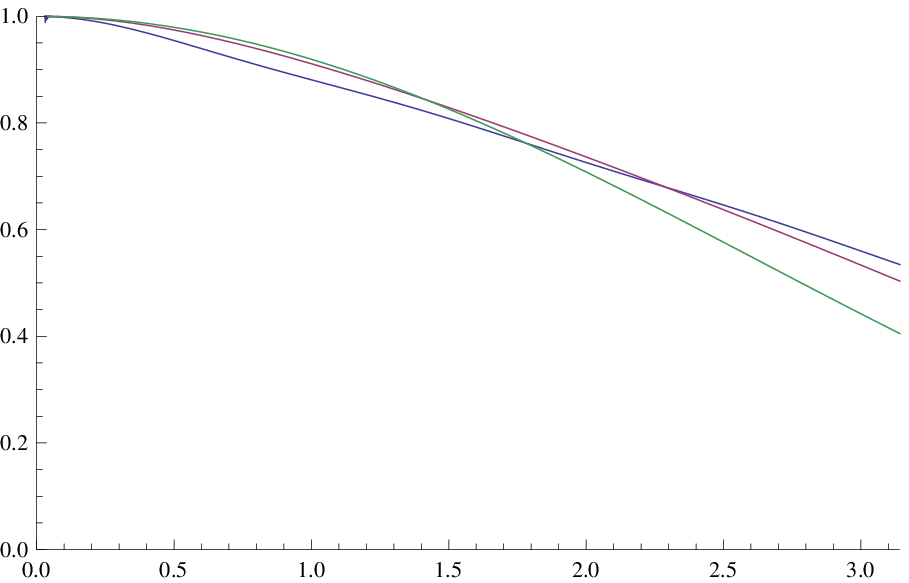}
   \end{center}
   \caption{The plot of the ratio $\f{\la \Psi_\ap |O(1)O(e^{-i\sigma})| \Psi_\ap\lb}{\la 0|O(1)O(e^{-i\sigma})|0\lb}$ at the time $t=0$ (left graph) and $t=\pi/2$ (right graph)
    as a function of $\sigma$. At $t=0$, the excitation is localized at the center of AdS. While at $t=\pi/2$, the excitation is spread throughout AdS. We chose $h_{\ap}=1/2$ and $h_{O}=\f{c}{24}$ and made order $n=4$ approximation. The blue, red and green curve corresponds to the value $\ep=0.35, 1$, and $\infty$. Note that when $\ep=\infty$ the left and right graphs do coincide, because the corresponding BTZ solution is static. We can also see the two point function increases under the time evolution when $0\leq t\leq \pi/2$ for finite $\ep$.}
   \label{fig:largec1}
\end{figure}

We showed the numerical plot of the two point function ratio (\ref{tworat}) for the state $|\Psi_\ap(t)\lb$ in Fig.\ref{fig:largec1} at the time $t=0$ and $t=\pi/2$. For $\epsilon=\infty$ we
find that they coincide and this is because $|\Psi_\ap(t)\lb$ is equal to the primary state $|\ap\lb$, which is static. Indeed we can confirm the result for $\epsilon=\infty$ agrees with the holographic result (\ref{btzgh}) from the BTZ black hole.\footnote{Note that since we are assuming that $h_O$ is the order $O(c)$, we do not expect the geodesic approximation (\ref{georg}), which assumes $1\ll h_O\ll c$,  is valid. However, for the heavy primary field corresponding to $\ep\to\infty$ limit, the gravity dual is known to be the BTZ, which is a solution to the vacuum Einstein equation. Owing to this special situation, we find this agreement between our CFT result and the geodesic one.}
 For general $\ep$, we find that the two point function gets increases under the time evolution when $0\leq t\leq \pi/2$. This time evolution gets more enhanced as $\ep$ becomes smaller.

\subsection{Time Evolution of $|\Psi_\ap(t)\lb$ and Bulk Causal Propagations}

\begin{figure}[ttt]
   \begin{center}
      \includegraphics[height=5cm]{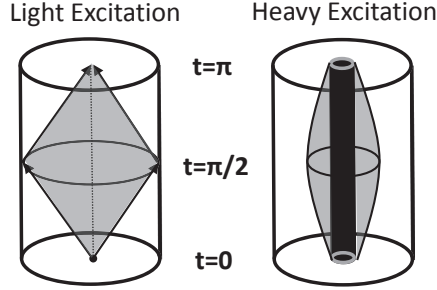}
   \end{center}
   \caption{The sketches of time evolution of the local excitation $|\Psi_\ap\lb$ in a global AdS$_3$.
   The gray regions describe the bulk scalar field excitation. In the white region we can ignore matter field contributions and the spacetime can be approximated by the BTZ black hole solution with the correct mass. The left picture corresponds to the light excitation $h_\ap\simeq 1/2$.
Since the dual bulk scalar field is light, we expect light like propagation of excitations.
The right picture describes the geometry for $h_\ap \gg 1$. Since the scalar field is very massive, the excitations does not reach the AdS boundary. Also the primary state includes in $|\Psi_\ap\lb$ gives a back reacted geometry (deficit angle spacetime) localized at the center of AdS.}\label{fig:timev}
\end{figure}

\begin{figure}[ttt]
   \begin{center}
     \includegraphics[height=4cm]{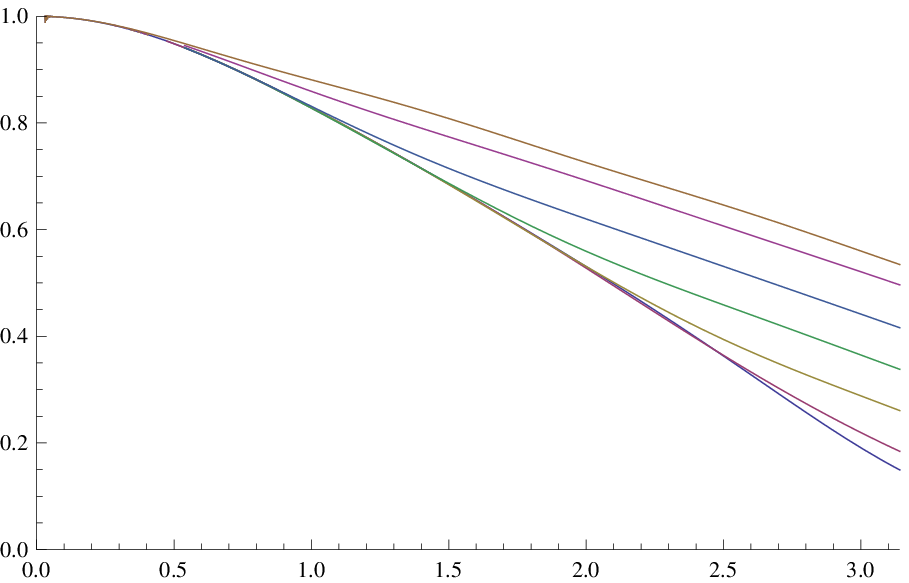}
     \hspace{1cm}
     \includegraphics[height=4cm]{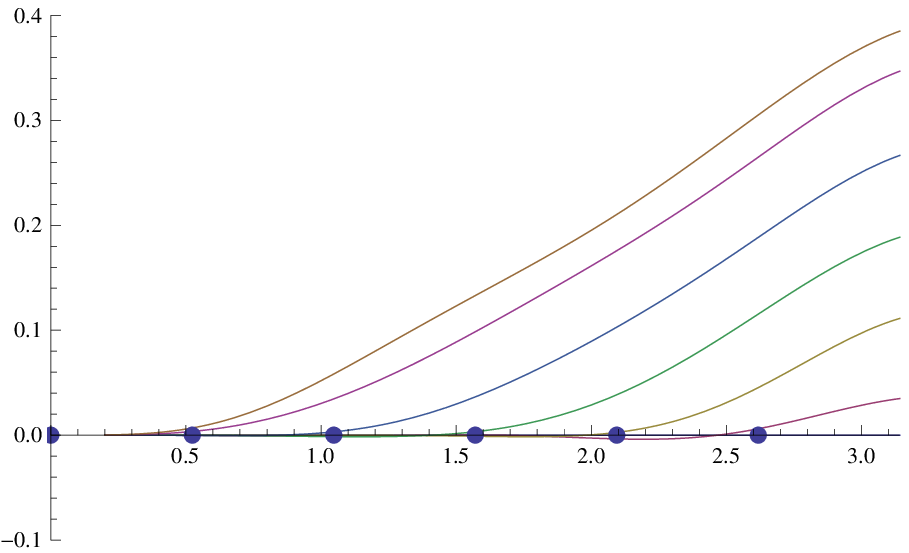}
   \end{center}
   \caption{The plot of the ratio $R(t)\equiv \f{\la \Psi_\ap(t) |O(1)O(e^{-i\sigma})| \Psi_\ap(t)\lb}{\la 0|O(1)O(e^{-i\sigma})|0\lb }$ (left graph) and the difference $R(t)-R(0)$
    (right graph) as a function of $\sigma$ for $t=0,\pi/6,\pi/3,\pi/2,2\pi/3,5\pi/6,\pi$
    (from the bottom to the top graph). At $t=0$, the excitation is localized at the center of AdS. While at $t=\pi/2$, the excitation is spread throughout AdS. We chose $h_{\ap}=1/2$ and $h_{O}=\f{c}{24}$ for $\ep=0.35$. We employed {$n=4$} approximation. In the right graph, we marked points $\sigma=0,\pi/6,\pi/3,\pi/2,2\pi/3,5\pi/6$ with blue dots, corresponding to the wave fronts as computed in (\ref{ww}).}\label{fig:largec3}
\end{figure}

Let us study the time evolution of the two point function more closely in the light of its holographic dual. Consider the gravity dual of the time evolution of the state $|\Psi_\ap(t)\lb$. The primary operator $O_\ap$ is dual to a bulk scalar field $\phi_\ap$ with mass $m_\ap=\Delta_\ap/R$.
At $t=0$, the origin $r=0$ of the global AdS$_3$ is locally excited. Note that we expect that the metric of the gravity dual even at $t=0$ will be modified everywhere due to the gravitational back reactions of the energetic excitation, though the scalar field expectation value $\la \phi_\ap\lb$ is non-vanishing only near the center of global AdS$_3$. Under the time evolution, the profile of scalar field $\phi_\ap$ gets oscillating and some amount of energy is emitted as gravitational waves. The total energy is given by (\ref{amd}). We expect that if $\epsilon$ is small and the dimension $\Delta_\ap$ of the operator is small enough such that $\Delta_{\Psi_\ap}\gg \Delta_\ap$, then large amount of excitations will spread at the speed of light, as sketched in the left picture of Fig.\ref{fig:timev}.
This is because the mass of scalar field is small and the energy of oscillation is very large. In this holographic picture, the geometry at time $t=t_0$ is perturbed from that of $t=0$ only in a region where the light rays from the origin can reach within the time period $t_0$.

Let us estimate this wave front by assuming the global AdS$_3$ metric (\ref{gads}). The null geodesic from $r=0$ at time $t=0$ reaches $r=r_0$ at
\be
t_0=\arctan \left(\f{r_0}{R}\right).
\ee
By using (\ref{btzgo}), the wave front at $t=t_0$ can be probed by the space-like geodesic which connects
$\sigma=0$ and $\sigma=\sigma_0$, where $\sigma_0$ is given by
\be
\cos\sigma_0=\f{r_0^2-R^2}{r_0^2+R^2}.
\ee
This is solved as
\be
\sigma_0=\pi-2t_0. \label{ww}
\ee
Note that we assumed the range $0\leq t_0 \leq \pi/2$ and $0 \leq \sigma_0\leq \pi$.

We showed in Fig.\ref{fig:largec3} the time evolution of two point function for the excited state $|\Psi_\ap(t)\lb$ when the mass of dual bulk scalar field saturates the BF bound i.e. $h_{\ap}=1/2$
(or $\Delta_\ap=1$) and therefore is lightest. At $t=0$ we find that the non-trivial suppression of two point function, which reflects the localized bulk excitation at the origin of AdS $r=0$. Under the time evolution, the suppression of two point function gets reduced for large enough values of $\sigma$ such that $\sigma\geq \pi-2t$ in a good approximation (see the right graph in Fig.\ref{fig:largec3}). Thus we conclude that this shows a light-like spread of the localized bulk excitation in AdS as expected. In this way, we observed that the time evolution of two point function shows bulk causal propagations. This is in contrast with the result in a free fermion CFT, where we do not find any causal propagations from the two point function as we show in Appendix A.

\begin{figure}[ttt]
   \begin{center}
     $\ap_O=0.5,~\ep=0.35$\\
     \includegraphics[height=3cm]{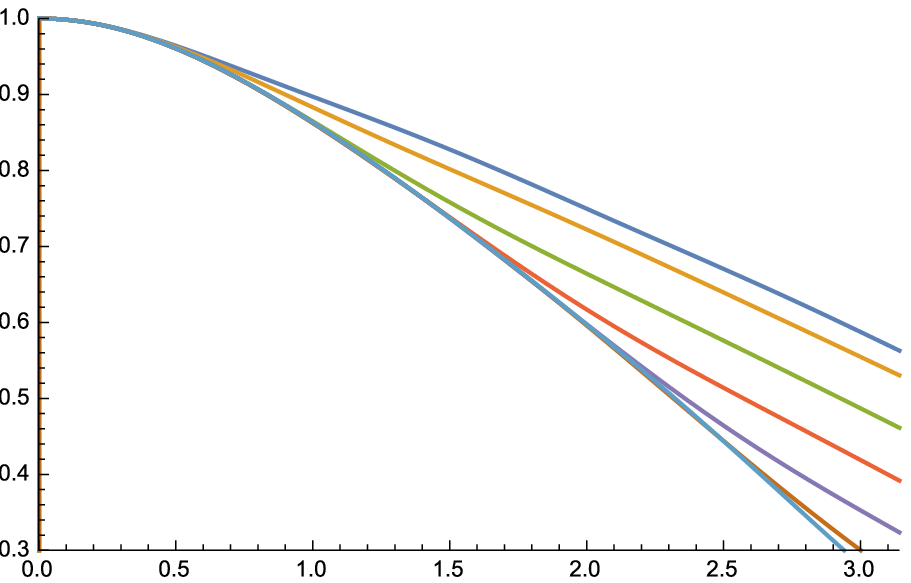}
      \hspace{1cm}
     \includegraphics[height=3cm]{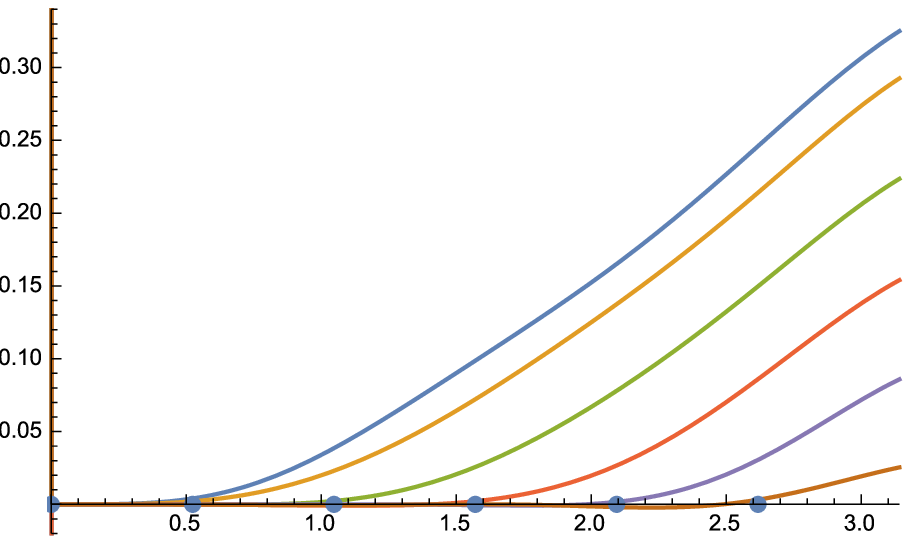}
      \hspace{2cm}\\
      $\ap_O=0.9,~\ep=0.35$\\
     \includegraphics[height=3cm]{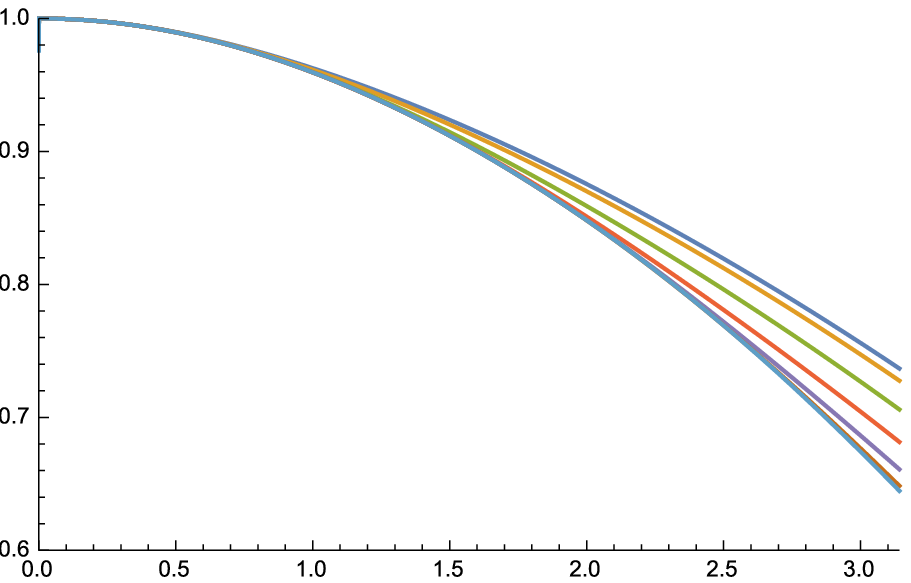}
      \hspace{1cm}
     \includegraphics[height=3cm]{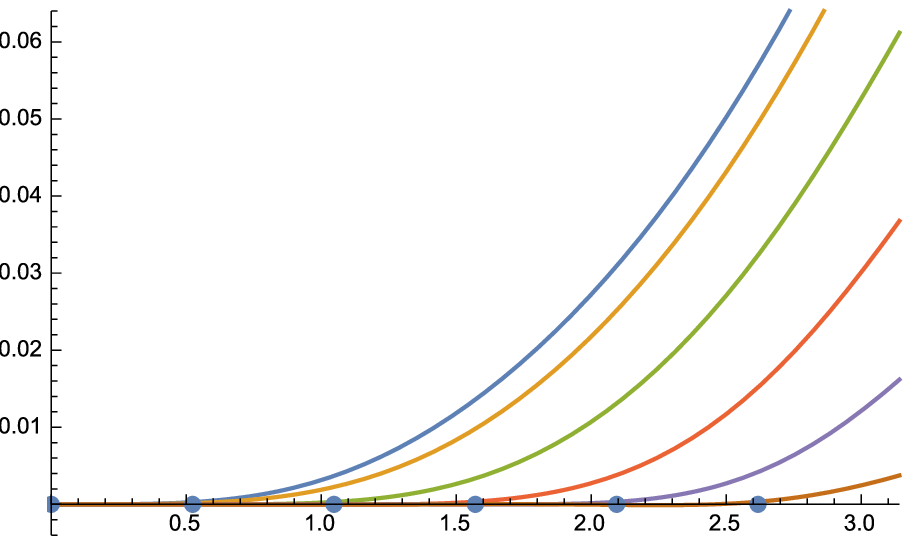}
\end{center}
   \caption{The two point function ratios for various $\ap_O=\sqrt{1-24h_{O}/c}$. The left graph \newline shows $R(t)\equiv\f{\la \Psi_\ap(t) |O(1)O(e^{-i\sigma})| \Psi_\ap(t)\lb}{\la 0|O(1)O(e^{-i\sigma})|0\lb }$ as a function of $\sigma$ for $t=0,\pi/6,\pi/3, \pi/2,2\pi/3,5\pi/6,\pi$ \newline from the bottom to top.
   The right graph describes the difference $R(t)-R(0)$ as a function of $\sigma$ for $t=0,\pi/6,\pi/3,\pi/2,2\pi/3,5\pi/6,\pi$. At $t=0$, the excitation is localized at the center of AdS. While at $t=\pi/2$, the excitation is spread throughout AdS. We chose $h_{\ap}=1/2$ and employed $n=4$ approximation. In the right graph, we marked points $\sigma=0,\pi/6,\pi/3,\pi/2,2\pi/3,5\pi/6$ with blue dots, corresponding to the light-like wavefront as computed in (\ref{ww}).} \label{fig:largeap}
\end{figure}

So far we assumed $\ap_O=0$ or equally $h_O=\f{c}{24}$ for the numerical plot. Now we show the behavior of two point functions for $\ap_O=0.5$ and $\ap_O=0.9$ in Fig.\ref{fig:largeap} when $h_\ap$ takes the same value $h_\ap=1/2$. We can confirm the causal propagations of excitations as in the previous section.

\subsection{Two Point Functions for Other values of $h_\ap$}

\begin{figure}[ttt]
   \begin{center}
    $h_\ap=1,~\ep=0.35$\\
     \includegraphics[height=3cm]{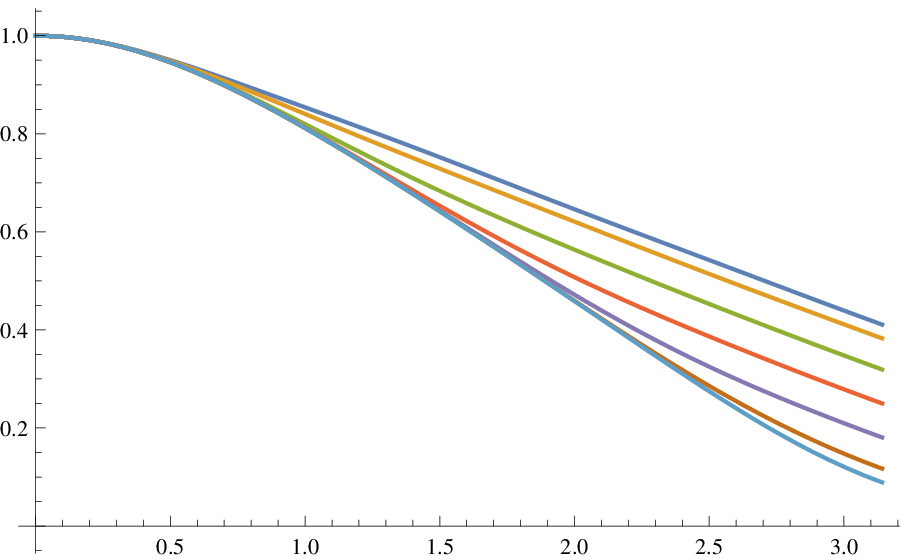}
      \hspace{1cm}
     \includegraphics[height=3cm]{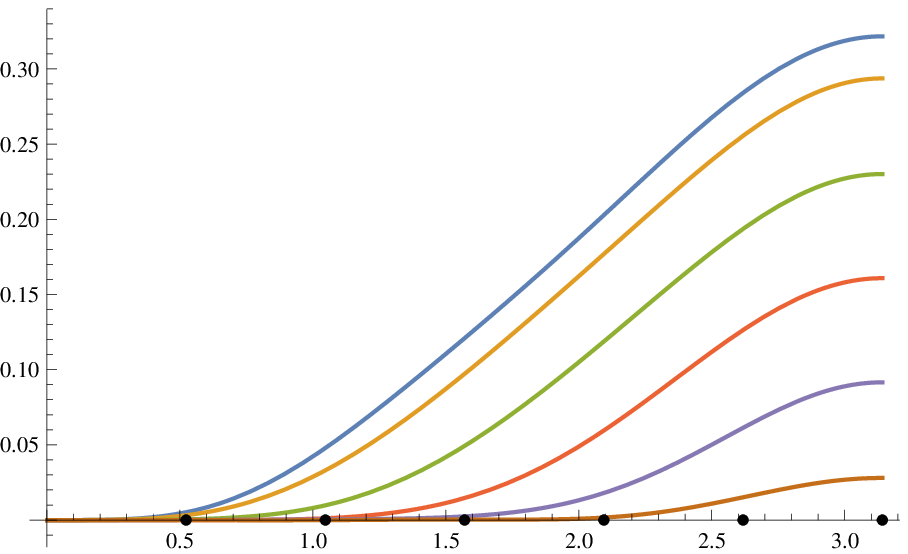}
      \hspace{2cm}\\
      $h_\ap=10,~\ep=0.35$\\
     \includegraphics[height=3cm]{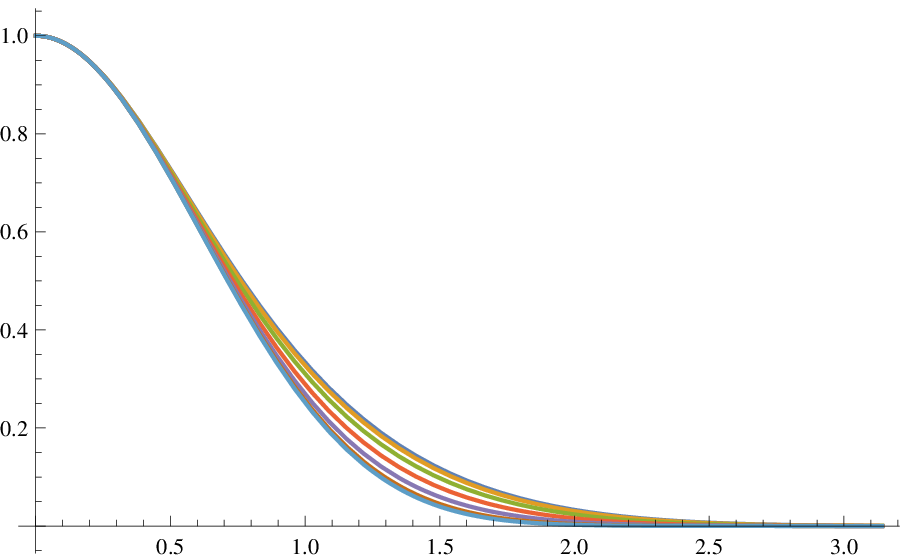}
      \hspace{1cm}
     \includegraphics[height=3cm]{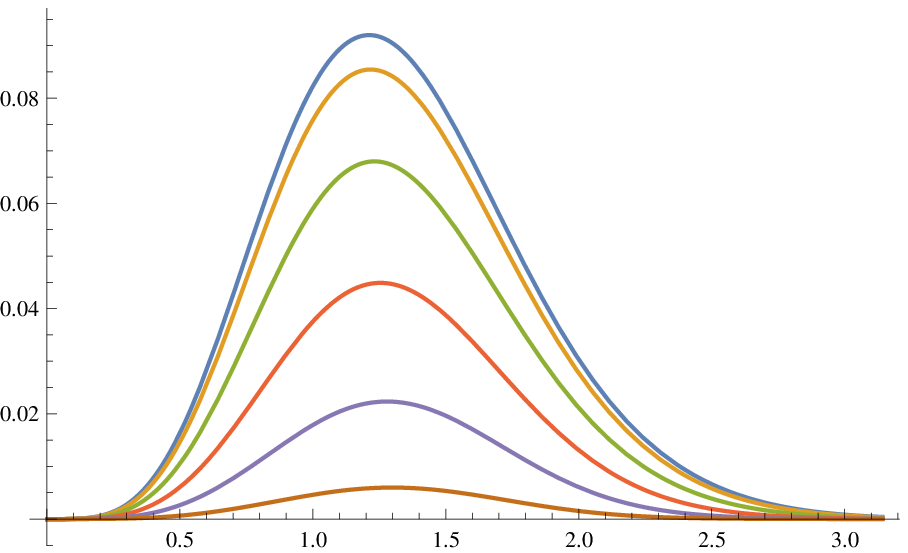}
\end{center}
   \caption{The two point function ratios for various $h_\ap$. The left graph shows $R(t)\equiv\f{\la \Psi_\ap(t) |O(1)O(e^{-i\sigma})| \Psi_\ap(t)\lb}{\la 0|O(1)O(e^{-i\sigma})|0\lb }$ as a function of $\sigma$ for $t=0,\pi/6,\pi/3,\pi/2,2\pi/3,5\pi/6,\pi$ from the bottom to top.
   The right graph describes the difference $R(t)-R(0)$ as a function of $\sigma$ for $t=0,\pi/6,\pi/3,\pi/2,2\pi/3,5\pi/6,\pi$ from the bottom to top. We chose $\ap=1/2$ and employed $n=4$ approximation. In the case $h_\alpha = 1$, we find approximate light-like wavefront as in $h_\alpha = 1/2$. At $h_{\alpha}=10$, we cannot observe light-like wavefront, because the bulk scalar field is too heavy.}\label{fig:largeh}
\end{figure}

Here we show the numerical plot of the two point function for $h_\ap=1$ and $h_\ap=10$. In the Fig.\ref{fig:largeh} we set $\epsilon=0.35$. In the case $h_\ap=1$ we find approximately a causal propagations of radiations as in the previous result for $h_\ap=1/2$. On the other hand, when $h_\ap=10$, the energy of the primary state $\Delta_{\alpha}$ dominates the total energy $\Delta_{\Psi_\ap}$ and the effect of radiations is relatively small since $\Delta_{\alpha}\gg 1/ \epsilon$ in our numerical computation. Thus we expect that the dual spacetime geometry consists of a static core dual to the state $|\ap\lb$ plus a small amount of radiations. Geodesics on the this back reacted geometry become very long in the region $\sigma \sim \pi $, and thus all the correlation functions of  $\sigma \sim \pi $ are suppressed. Also there is a potential barrier as we approach the AdS boundary due to the large mass of bulk scalar field. These qualitatively explain the behavior of our results (refer to the right picture in Fig.\ref{fig:timev}).

\section{Analysis for $h_\ap \ll c$ and $h_O\ll c$ in Holographic CFT}

So far we assume $h_\ap=O(1)$ and $h_O=O(c)$. In this case, though we want to employ the operator $O$ as a probe, its conformal dimension is the same order of the central charge and we cannot neglect the back reaction by the operator $O$. This motivates us to study another region: $h_O\ll c$. Motivated by this, we would like to focus on the region $\f{h_\ap}{c}\ll 1$ and $\f{h_O}{c}\ll 1$ in this section.

\begin{figure}[p]
   \begin{center}
        $h_\ap=1/2,\ h_O=100,\ c=10000, \ep=0.5$\\
      \includegraphics[height=3cm]{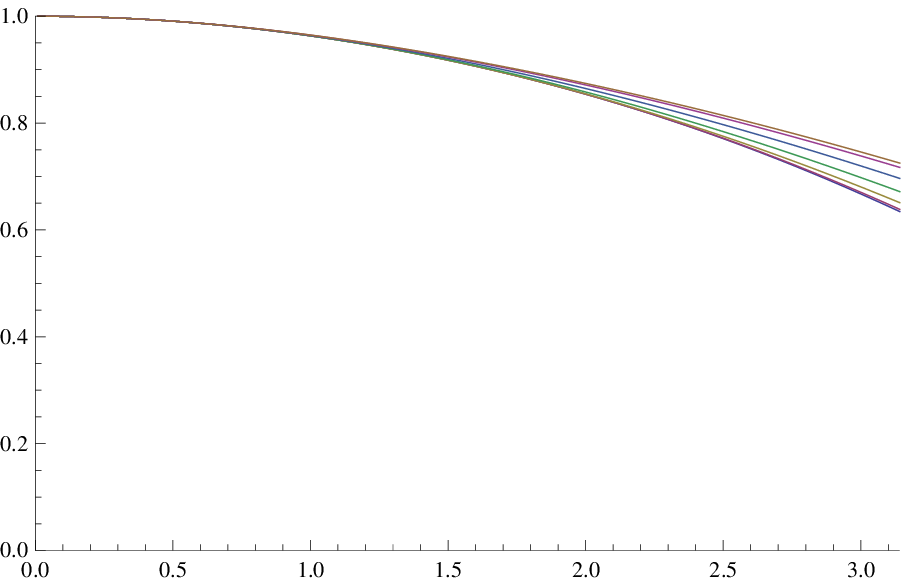}
      \hspace{1cm}
     \includegraphics[height=3cm]{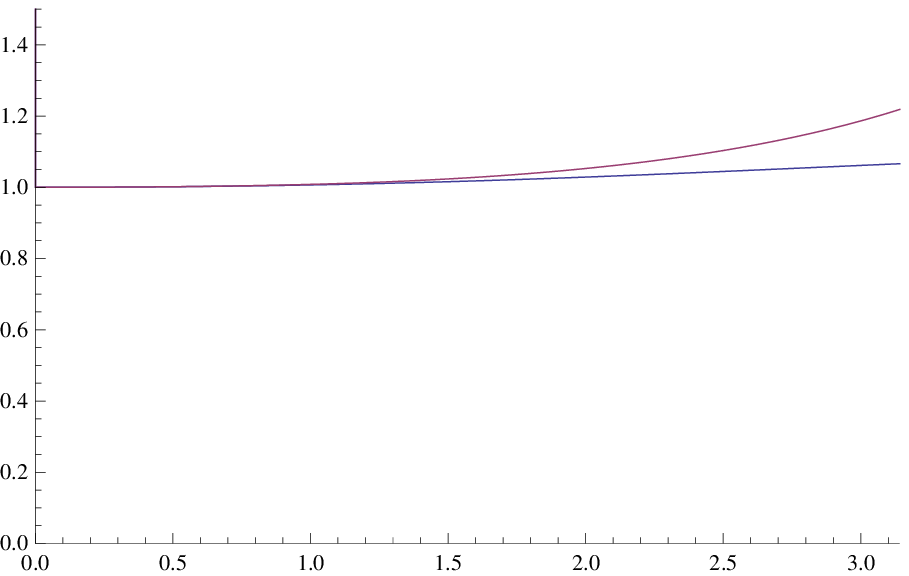}
     \hspace{2cm}\\
      \vspace{5mm}
       $h_\ap=10,\ h_O=30,\ c=10000, \ep=0.5$\\
      \includegraphics[height=3cm]{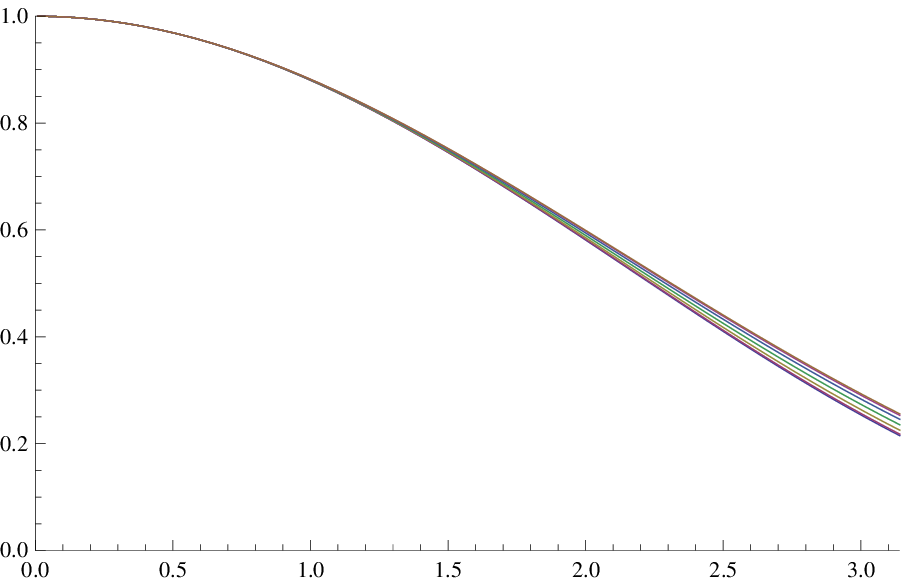}
      \hspace{1cm}
     \includegraphics[height=3cm]{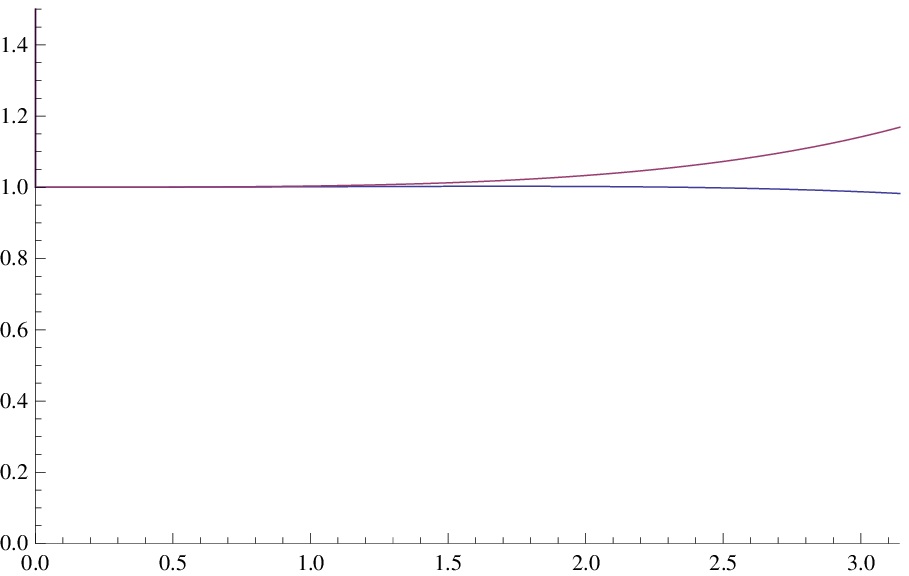}
   \end{center}
   \caption{The profiles of two point functions in the two choices of the parameters:
   $(h_\ap,h_O,c)=(1/2,100,10000)$ (upper) and $(h_\ap,h_O,c)=(10,30,10000)$ (lower). The left graphs show the quantity $R(t)\equiv \f{\la \Psi_\ap(t) |O(1)O(e^{-i\sigma})| \Psi_\ap(t)\lb}{\la 0|O(1)O(e^{-i\sigma})|0\lb }$ as a function of $\sigma$ for $t=0,\pi/6,\pi/3,\pi/2,2\pi/3,5\pi/6,\pi$ from the bottom to top. The right graphs show this two point function $R(t)$ divided by the holographic result from the geodesic length in BTZ black hole with mass corresponding to $\Delta_{\Psi_\ap}$. The blue and red curve correspond to this ratio at $t=0$ and $t=\pi/2$, respectively. In each case we set $\ep=0.5$ and employed the $n=2$ order approximation. In both graphs, we find that the time dependence is weak compared to that in Fig.\ref{fig:largec3}. We find that at $t=0$ the profile is very close to that obtained from a BTZ, while at $t=\pi/2$ they differs largely especially in the region where $\sigma$ is large. This shows that the spacetime at $t=0$ can be approximated well by the BTZ  solution, on the other hand, the spacetime at $t\neq0$ is a non-trivial solution of the Einstein equation coupled to the scalar field which is quite different from the BTZ solution. The behavior of these graphs mentioned above also explains the fact that even at $t\neq0$ the geometry outside the excitation should be a BTZ black hole with the mass corresponding to $\Delta_{\Psi_\ap}$. }\label{fig:sq}
\end{figure}

We can evaluate the two point function by simply plugging the formula (\ref{hh}) in (\ref{jwje}) and compute the two point function numerically. We show the behavior of two point functions as functions of $\sigma$ for
$(h_\ap,h_O,c)=(1/2,100,10000)$ and $(h_\ap,h_O,c)=(10,30,10000)$ in Fig.\ref{fig:sq}. In both cases, we find that the time dependence is not strong compared with that in Fig.\ref{fig:largec3}. This is because in the current case, $h_O$ is much smaller than $c$ and therefore the probe operator $O$ cannot amplify the back reaction due to the excited state $|\Psi_\ap\lb$.

In the first case $(h_\ap,h_O,c)=(1/2,100,10000)$, the result is similar to the result for $(h_\ap,\ap_O)=(1/2,0.9)$ in Fig.\ref{fig:largeap}. On the other hand, in the second one $(h_\ap,h_O,c)=(10,30,10000)$, we find that the time-dependence is more suppressed and its holographic interpretation is the same as that depicted in the right picture of Fig.\ref{fig:timev}.

In Fig.\ref{fig:sq}, we also plotted the ratios of the CFT two point functions against that from the BTZ geodesic (\ref{btzgh}) in Fig.\ref{fig:largec3} at the times $t=0$ and $t=\pi/2$. From this we learn that at $t=0$ the profile is very close to that obtained from a BTZ black hole, while at $t=\pi/2$ they differs largely. This shows that the spacetime metric at $t=0$ is well approximated by the BTZ black hole solution. This supports our claim in Fig.\ref{fig:timev}. The region outside the causal evolution of the localized excitation, which is the white region in Fig.\ref{fig:timev}, can be described by a solution of the vacuum Einstein equation and therefore it should be a BTZ black hole with the mass corresponding to $\Delta_{\Psi_\ap}$.  The behavior of the right plots in Fig.\ref{fig:sq} in the region where $\sigma$ is not large shows this expectation is indeed right. The region inside should be described by a non-trivial solution of Einstein equation coupled to the scalar field matter. In order to know the interior geometry exactly, we need a further analysis for the Einstein-scalar system. However, in each plot we can see that  the suppression of the correlation function gets reduced in the region where $\sigma$ is large, thus we expect that the deformation of the interior geometry from the pure AdS is smaller than that of the exterior geometry, i.e. a BTZ black hole spacetime with the mass $\Delta_{\Psi_\ap}$.

\section{Conclusions}

In this article, we studied time evolutions of a class of excited states $|\Psi_{\ap}\lb$ in two dimensional holographic CFTs, which are expected to be dual to localized excitations in the bulk AdS$_3$, in order to understand the bulk locality and causality purely from CFTs. We computed two point functions of heavy primary operators to probe back-reactions in the bulk geometry from the localized bulk excitations. We found that excited states dual to a scalar field with a mass close to the BF bound show the time evolutions which are dual to light-like spread of the localized excitations in the bulk AdS. On the other hand, those dual to a very massive scalar field show weak time evolutions localized in the middle of AdS space, which is interpreted as non-relativistic excitations of the scalar field in the AdS. In both cases, we found that outside the light cone, the bulk geometry can be well approximated by BTZ black hole with the mass $\Delta_{\Psi_\alpha}$. We also found that the short distance behavior of the two point function satisfies the relation analogous to the first law of entanglement entropy. Note that there have been many works which 
employ the large $c$ approximation in 2d holographic CFTs to extract bulk classical gravity, the present paper has a unique feature that it focuses on an excitation which is very elementary from the bulk gravity i.e. local excitation in a bulk point. Our study in this paper revealed the emergence of bulk light cone structure purely from CFT computations. 

An important future problem is to well understand the geometry inside the light cone, where the scalar field is non-trivially excited. To explore the interior of the light cone, we need a careful analysis of Einstein-scalar theory. We note that our excited state $|\Psi_{\ap}\lb$ can have arbitrary large energy and has an exact periodicity $\pi$ in time direction. It is intriguing to find a corresponding gravity solution with such a periodicity, because generically localized matter in gravity theory does not have such periodicity and tend to form a black hole finally \cite{BR,Lopez}.

\section*{Acknowledgements}

  We are grateful to Tokiro Numasawa for useful discussions. KG and MM are supported by JSPS fellowship. TT is supported by the Simons Foundation and JSPS Grant-in-Aid for Scientific Research (A) No.16H02182. TT is also supported by World Premier International Research Center Initiative (WPI Initiative) from the Japan Ministry of Education, Culture, Sports, Science and Technology (MEXT).

\newpage
\appendix

\section{Two Point Functions for $|\Psi_\ap\lb$ in Free Fermion CFT}

In this section, as an toy analytical example, we would like to evaluate the two point function (\ref{jwje}) for the $c=1$ CFT (free Dirac fermion or free scalar) by using (\ref{dedf}). We choose $O=e^{i\beta\phi}$. We can
express (\ref{jwje}) in the following form:
\ba
&& \la \Psi_\ap|O(1)O(e^{-i\sigma})|\Psi_\ap\lb \no
&& ={\cal N}\cdot|1-e^{-i\sigma}|^{-2\beta^2}\cdot\lim_{x, u\to 0} {\cal D}_x{\cal D}_u \left[|1-xu|^{-2}
|1-\eta|^{-2\beta}\right] \no
&&={\cal N}\cdot |1-e^{-i\sigma}|^{-2\beta^2}\cdot\lim_{x, u\to 0} {\cal D}_x{\cal D}_u \left[I(x,u,\bar{x},\bar{u},\sigma)\right],
\label{wwle}
\ea
where we defined
\ba
&& I(x,u,\bar{x},\bar{u},\sigma) \no
&& =\left(\sum_{k=0}^\infty x^k u^k\right) \left(\sum_{\bar{k}=0}^\infty \bar{x}^{\bar{k}} \bar{u}^{\bar{k}} \right)\left(\sum_{l=0}^\infty c_lx^l\right)\left(\sum_{\bar{l}=0}^\infty \bar{c}_{\bar{l}}\bar{x}^{\bar{l}}\right)\left(\sum_{m=0}^\infty d_m u^m\right)\left(\sum_{\bar{m}=0}^\infty \bar{d}_{\bar{m}}\bar{u}^{\bar{m}}\right).\no
\ea

The actions of ${\cal D}_{x,u}$ (\ref{ddxu}) and the limits $x,u\to 0$ tell us that only the terms in $I(x,u,\bar{x},\bar{u},\sigma)$ which satisfy
\be
p=k+l=\bar{k}+\bar{l},\ \ \ \ q=k+m=\bar{k}+\bar{m},
\ee
contribute to the final result.

Then it is convenient to decompose the summation into three parts $p>q$, $q<p$ and $p=q$. In this way we obtain the following evaluation:
\ba
&& \la \Psi_\ap(t)|O(1)O(e^{-i\sigma})|\Psi_\ap(t)\lb \cdot{\cal N}^{-1} \cdot |1-e^{-i\sigma}|^{2\beta^2} \no
&& =\sum_{p=0}^\infty \sum_{q=0}^{p-1}e^{2i(q-p)t}(-1)^{p+q}e^{-\ep(p+q)}\left|\sum_{m=0}^q c_{p-q+m}d_m\right|^2, \no
&& \ \ \ +\sum_{p=0}^\infty \sum_{q=p+1}^{\infty}e^{2i(q-p)t}(-1)^{p+q}e^{-\ep(p+q)}\left|\sum_{l=0}^p c_l d_{q-p+l}\right|^2 \no
&& \ \ \ +\sum_{p=0}^\infty e^{-2\ep p}\left|\sum_{m=0}^p c_{m}d_m\right|^2.
\ea

Now we focus on the special case $\beta=1$.  In this case we find
\be
c_0=d_0=1,\ \ \ \  c_{l\geq 1}=(1-e^{-i\sigma})e^{il\sigma},\ \ \ \ d_{l\geq 1}=1-e^{-i\sigma}.
\ee
Thus we can evaluate each of summations as follows:
\ba
&& \sum_{m=0}^q c_{p-q+m}d_m=\sum_{l=0}^p c_l d_{q-p+l}=(1-e^{-i\sigma})e^{ip\sigma}, \no
&& \sum_{m=0}^p c_{m}d_m=(1-e^{-i\sigma})e^{ip\sigma}+e^{-i\sigma}.
\ea
Finally we can compute the two point function (for $O_B=e^{i\phi}$) as follows
\ba
&& \la \Psi_\ap(t)|O(1)O(e^{-i\sigma})|\Psi_\ap(t)\lb \no
&& ={\cal N}\cdot\Bigl[ |1-e^{-i\sigma}|^{-2}\cdot \sum_{p=0}^\infty e^{-2p\ep}\left(1+(1-e^{i\sigma})
e^{-i(p+1)\sigma}+(1-e^{-i\sigma})e^{i(p+1)\sigma}\right) \no
&& \ \ \ \ \ \ \ \ \ \ \ \ +\sum_{p=0}^\infty\sum_{q=0}^\infty
e^{2i(q-p)t}(-1)^{p+q} e^{-\ep(p+q)}\Bigr], \no
&&= |1-e^{-i\sigma}|^{-2}\cdot \left[1+\left(\f{e^{-i\sigma}-1}{1-e^{-2\ep-i\sigma}}
+\f{e^{i\sigma}-1}{1-e^{-2\ep+i\sigma}}\right)(1-e^{-2\ep})\right]\no
&& \ \ \ \ +\f{1-e^{-2\ep}}
{(1-e^{2it}e^{-\ep})(1-e^{-2it}e^{-\ep})}, \no
\label{sumrq}
\ea

\begin{figure}[ttt]
   \begin{center}
     \includegraphics[height=4cm]{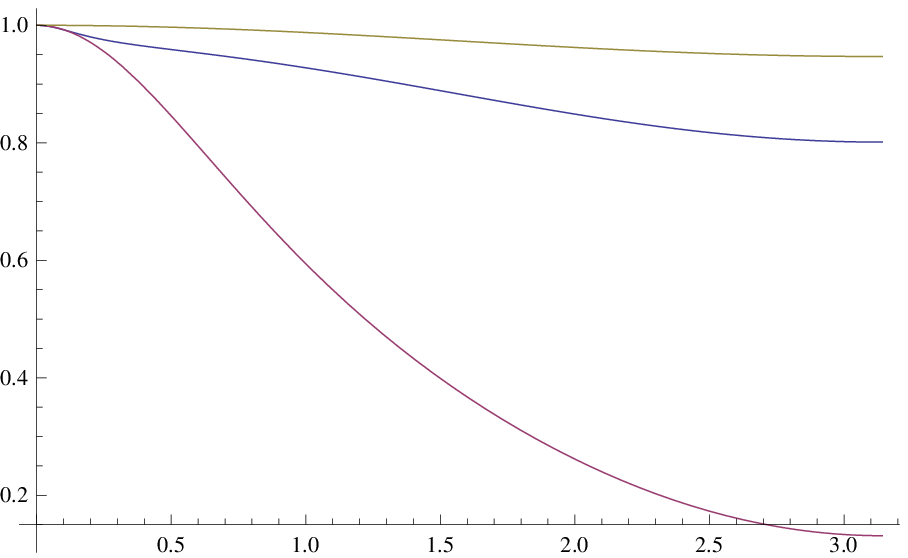}
      \hspace{1cm}
     \includegraphics[height=4cm]{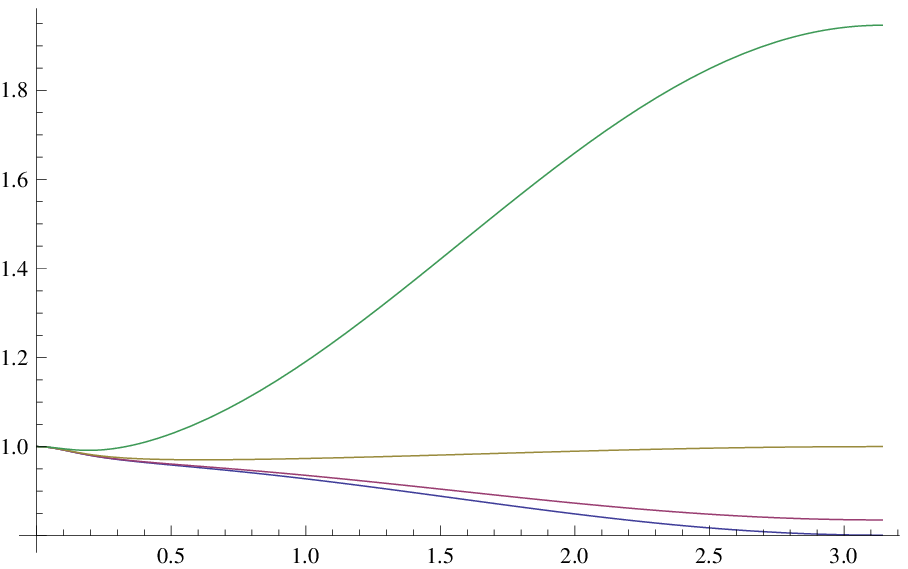}
   \end{center}
   \caption{The left graph shows a plot of the ratio $\f{\la \Psi_\ap |O(1)O(e^{-i\sigma})| \Psi_\ap\lb}{\la 0|O(1)O(e^{-i\sigma})|0\lb }$ as a function of $\sigma$ for $\beta=1$, for free fermion CFT. The blue, red and yellow curve corresponds to the value $\ep=0.1, 0.5$ and $5$. The right plot describes the same ratio at $t=0$ (blue), $t=\pi/4$ (red), $t=\pi/2$ (yellow), $t=3\pi/4$ (green) for $\ep=0.1$.}
   \label{fig:free1}
\end{figure}

Thus at $t=0$ we just have in the limit $\ep\to 0$
\ba
\la \Psi_\ap|O(1)O(e^{-i\sigma})|\Psi_\ap\lb \simeq |1-e^{-i\sigma}|^{-2}.
\ea
At the same time, we should notice that this is a special example because if we just consider a primary state $|\ap\lb$ we find the same result (this is true for any $\beta$):
\be
\la \ap|O(1)O(e^{-i\sigma})|\ap\lb =|1-e^{-i\sigma}|^{-2}.
\ee
This may be surprising as we cannot see the difference between the primary state $|\ap\lb$ and
our new state $|\Psi_\ap\lb$. Moreover, they are even the same as the two point function for the vacuum state $|0\lb$. However we would like to interpret this as an artifact of free CFTs.
For interacting CFTs this does not happen in general as we will see in the next section.
For non-zero $\ep$, we find a non-trivial behavior as in Fig.\ref{fig:free1}. In both the limit $\ep\to 0$ and  $\ep\to \infty$ the results coincide with the vacuum two point function $\la 0|OO|0\lb$.

We would also like to mention that this result does not satisfy the short distance behavior
(\ref{firstjg}), which means that the free fermion CFT is an exception. A similar result is known for entanglement entropy as found in \cite{ABS}.

Now, at $t=\pi/2$ we obtain
\ba
&& \la \Psi_\ap(\pi/2)|O(1)O(e^{-i\sigma})|\Psi_\ap(\pi/2)\lb  \no
&& = |1-e^{-i\sigma}|^{-2}\cdot \left[1+\left(\f{e^{-i\sigma}-1}{1-e^{-2\ep-i\sigma}}
+\f{e^{i\sigma}-1}{1-e^{-2\ep+i\sigma}}\right)(1-e^{-2\ep})\right]+\f{1-e^{-2\ep}}{(1-e^{-\ep})^2}, \no
\ea
where only the last term is modified. If we take the limit $\ep\to 0$ we get
\be
\la \Psi_\ap(\pi/2)|O(1)O(e^{-i\sigma})|\Psi_\ap(\pi/2)\lb \simeq |1-e^{-i\sigma}|^{-2}+\f{2}{\ep}.
\ee
This shows that the time evolution gives a larger correction to the two point function.

We plotted the behavior of the two point function in Fig.\ref{fig:free1} for several
$\ep$ and $t$ at $\beta=1$. Its time evolution is indeed different from what we expect from the causal propagation in AdS space. This is consistent with the fact that there is no classical gravity dual to a free fermion CFT.

\begin{thebibliography}{}




\bibitem{Ma}
  J.~M.~Maldacena,
  ``The large N limit of superconformal field theories and supergravity,''
  Adv.\ Theor.\ Math.\ Phys.\  {\bf 2} (1998) 231
  [Int.\ J.\ Theor.\ Phys.\  {\bf 38} (1999) 1113]
  [arXiv:hep-th/9711200];


\bibitem{GKP}
S.~S.~Gubser, I.~R.~Klebanov and A.~M.~Polyakov,
  ``Gauge theory correlators from non-critical string theory,''
  Phys.\ Lett.\ B {\bf 428}, 105 (1998);
E.~Witten,
  ``Anti-de Sitter space and holography,''
  Adv.\ Theor.\ Math.\ Phys.\  {\bf 2}, 253 (1998).


\bibitem{HKLL}
  A.~Hamilton, D.~N.~Kabat, G.~Lifschytz and D.~A.~Lowe,
  ``Local bulk operators in AdS/CFT: A Boundary view of horizons and locality,''  Phys.\ Rev.\ D {\bf 73} (2006) 086003  doi:10.1103/PhysRevD.73.086003  [hep-th/0506118];  
  ``Holographic representation of local bulk operators,''  Phys.\ Rev.\ D {\bf 74} (2006) 066009  doi:10.1103/PhysRevD.74.066009  [hep-th/0606141].  



\bibitem{HPPS}
  I.~Heemskerk, J.~Penedones, J.~Polchinski and J.~Sully,
  ``Holography from Conformal Field Theory,''  JHEP {\bf 0910} (2009) 079  doi:10.1088/1126-6708/2009/10/079  [arXiv:0907.0151 [hep-th]].  




\bibitem{ADH}
  A.~Almheiri, X.~Dong and D.~Harlow,
  ``Bulk Locality and Quantum Error Correction in AdS/CFT,''  JHEP {\bf 1504} (2015) 163  doi:10.1007/JHEP04(2015)163  [arXiv:1411.7041 [hep-th]].  





\bibitem{KL}
  D.~Kabat and G.~Lifschytz,
  ``Bulk equations of motion from CFT correlators,''  JHEP {\bf 1509} (2015) 059  doi:10.1007/JHEP09(2015)059  [arXiv:1505.03755 [hep-th]];  
 D.~Kabat and G.~Lifschytz,
  ``Locality, bulk equations of motion and the conformal bootstrap,''  arXiv:1603.06800 [hep-th].  



\bibitem{MSZ}
J.~Maldacena, D.~Simmons-Duffin and A.~Zhiboedov,
``Looking for a bulk point,''  arXiv:1509.03612 [hep-th].  




\bibitem{MNSTW}
  M.~Miyaji, T.~Numasawa, N.~Shiba, T.~Takayanagi and K.~Watanabe,
  ``Continuous Multiscale Entanglement Renormalization Ansatz as Holographic Surface-State Correspondence,''
  Phys.\ Rev.\ Lett.\  {\bf 115} (2015) 17,  171602
  doi:10.1103/PhysRevLett.115.171602
  [arXiv:1506.01353 [hep-th]].




\bibitem{Is}
  N.~Ishibashi,
  ``The Boundary and Crosscap States in Conformal Field Theories,''  Mod.\ Phys.\ Lett.\ A {\bf 4} (1989) 251.  




\bibitem{Ve}
  H.~Verlinde,
  ``Poking Holes in AdS/CFT: Bulk Fields from Boundary States,''
  arXiv:1505.05069 [hep-th].



\bibitem{NO}
  Y.~Nakayama and H.~Ooguri,
  ``Bulk Locality and Boundary Creating Operators,''
  JHEP {\bf 1510} (2015) 114
  doi:10.1007/JHEP10(2015)114
  [arXiv:1507.04130 [hep-th]].



\bibitem{NaOo}
  Y.~Nakayama and H.~Ooguri,
  ``Bulk Local States and Crosscaps in Holographic CFT,''  arXiv:1605.00334 [hep-th].  




\bibitem{Wang}
  Z.~L.~Wang,
  ``Bulk Local Operators, Conformal Descendants and Radial Quantization,''  arXiv:1507.05550 [hep-th].  




\bibitem{Ha}
  T.~Hartman,
  ``Entanglement Entropy at Large Central Charge,''  arXiv:1303.6955 [hep-th];  



\bibitem{FKW}
  A.~L.~Fitzpatrick, J.~Kaplan and M.~T.~Walters,
  ``Universality of Long-Distance AdS Physics from the CFT Bootstrap,''  JHEP {\bf 1408} (2014) 145  doi:10.1007/JHEP08(2014)145  [arXiv:1403.6829 [hep-th]].  



\bibitem{HKS}
  T.~Hartman, C.~A.~Keller and B.~Stoica,
  ``Universal Spectrum of 2d Conformal Field Theory in the Large c Limit,''  JHEP {\bf 1409} (2014) 118  doi:10.1007/JHEP09(2014)118  [arXiv:1405.5137 [hep-th]].  



\bibitem{RT}
S.~Ryu and T.~Takayanagi,
  ``Holographic derivation of entanglement entropy from AdS/CFT,''  Phys.\ Rev.\ Lett.\  {\bf 96} (2006) 181602  doi:10.1103/PhysRevLett.96.181602  [hep-th/0603001];  
 V.~E.~Hubeny, M.~Rangamani and T.~Takayanagi,
  ``A Covariant holographic entanglement entropy proposal,''  JHEP {\bf 0707} (2007) 062  doi:10.1088/1126-6708/2007/07/062  [arXiv:0705.0016 [hep-th]].  

\bibitem{cftbook}
P.~Di Francesco, P.~Mathieu and D.~Senechal,
  ``Conformal Field Theory,''  doi:10.1007/978-1-4612-2256-9  


\bibitem{BF}
P.~Breitenlohner and D.~Z.~Freedman,
``Positive energy in anti-de Sitter backgrounds and gauged extended supergravity,''
Phys.\ Lett.\ B {\bf 115} (1982) 197 doi:10.1016/0370-2693(82)90643-8\\



\bibitem{FKWt}
  A.~L.~Fitzpatrick, J.~Kaplan and M.~T.~Walters,
  ``Virasoro Conformal Blocks and Thermality from Classical Background Fields,''  JHEP {\bf 1511} (2015) 200  doi:10.1007/JHEP11(2015)200  [arXiv:1501.05315 [hep-th]].  



\bibitem{BNTU}
  J.~Bhattacharya, M.~Nozaki, T.~Takayanagi and T.~Ugajin,
  ``Thermodynamical Property of Entanglement Entropy for Excited States,''  Phys.\ Rev.\ Lett.\  {\bf 110} (2013) no.9,  091602  doi:10.1103/PhysRevLett.110.091602  [arXiv:1212.1164].  

\bibitem{ABS}
  F.~C.~Alcaraz, M.~I.~Berganza and G.~Sierra,
  ``Entanglement of low-energy excitations in Conformal Field Theory,''  Phys.\ Rev.\ Lett.\  {\bf 106} (2011) 201601  doi:10.1103/PhysRevLett.106.201601  [arXiv:1101.2881 [cond-mat.stat-mech]].  




\bibitem{BTZ}
 M.~Banados, C.~Teitelboim and J.~Zanelli,
  ``The Black hole in three-dimensional space-time,''  Phys.\ Rev.\ Lett.\  {\bf 69} (1992) 1849  doi:10.1103/PhysRevLett.69.1849  [hep-th/9204099].  



\bibitem{BR}
  P.~Bizon and A.~Rostworowski,
  ``On weakly turbulent instability of anti-de Sitter space,''  Phys.\ Rev.\ Lett.\  {\bf 107} (2011) 031102  doi:10.1103/PhysRevLett.107.031102  [arXiv:1104.3702 [gr-qc]].  

\bibitem{Lopez}
  J.~Abajo-Arrastia, E.~da Silva, E.~Lopez, J.~Mas and A.~Serantes,
  ``Holographic Relaxation of Finite Size Isolated Quantum Systems,''  JHEP {\bf 1405} (2014) 126  doi:10.1007/JHEP05(2014)126  [arXiv:1403.2632 [hep-th]].  





\end{thebibliography}
\end{document}